\documentstyle[prb,aps,epsfig]{revtex}

\begin{document}
\draft

\title{Experimental Critical Current Patterns in Josephson Junction
Ladders}

\author{P. Binder, P. Caputo, M. V. Fistul, and A. V. Ustinov}
\address{Physikalisches Institut III, Universit\"at Erlangen,
  E.-Rommel-Stra\ss e 1, D-91058 Erlangen, Germany}

\author{G. Filatrella}
\address{INFM Unit Salerno and Science Faculty, University of Sannio,
  Via Port'Arsa 11, I-82100 Benevento, Italy }

\date{\today}

\wideabs{ 

\maketitle

\begin{abstract}
  We present an experimental and theoretical study of the magnetic
  field dependence of the critical current of Josephson junction
  ladders. At variance with the well-known case of a one-dimensional
  (1D) parallel array of Josephson junctions the magnetic field
  patterns display a single minimum even for very low values of the
  self-inductance parameter $\beta_{\rm L}$.  Experiments performed
  changing both the geometrical value of the inductance and the
  critical current of the junctions show a good agreement with
  numerical simulations.  We argue that the observed magnetic field
  patterns are due to a peculiar mapping between the isotropic
  Josephson ladder and the 1D parallel array with the self-inductance
  parameter $\beta_{\rm L}^{\rm \/eff}=\beta_{\rm L}+2$.
\end{abstract}


}

\section{Introduction}
\begin{figure}
\vspace{5pt}
\centerline{\epsfig{file=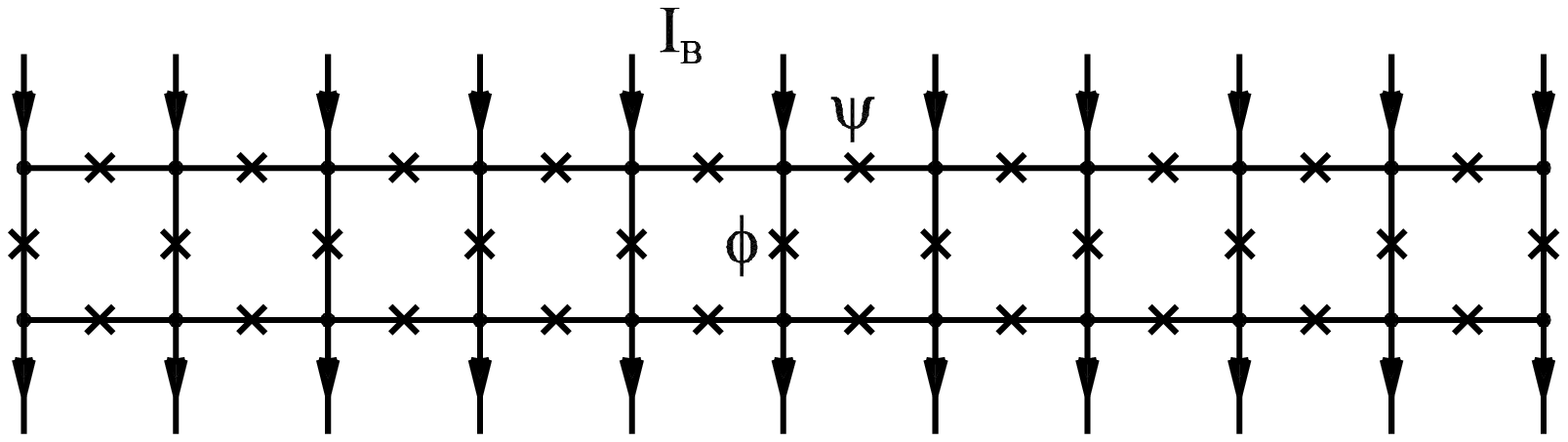,width=3.4in}}
\noindent(a)\vspace{1.5mm}
\centerline{\epsfig{file=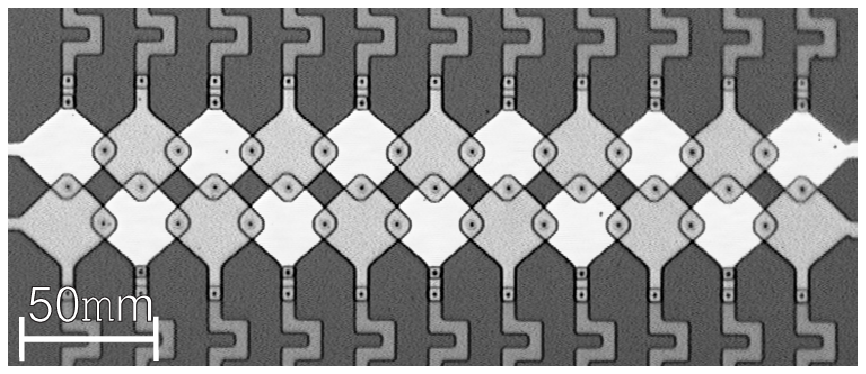,width=3.4in}}
\noindent(b)
\vspace{5pt}
\caption{(a) The electrical scheme of a Josephson junction ladder;
  crosses ($\times$) indicate Josephson junctions. (b) Optical image of
  one of studied samples.}
\label{scheme}
\end{figure}

A Josephson junction ladder is an array of coupled superconducting
loops containing small Josephson junctions as shown schematically in
Fig.~\ref{scheme}(a).  In the past years, ladders of Josephson junctions
have attracted considerable interest for a number of reasons. On one
hand, more complex systems such as two-dimensional arrays of Josephson
junctions can be viewed as elementary ladders coupled to each other
\cite{filatrella95,basler}. On the other hand, the ladders are an
ideal ground for experimental and theoretical investigations of
discrete nonlinear
entities~\cite{bonilla91,ustinov93,watanabe95,caputo98}, such as
breathers~\cite{sfcrw98,fmmfa96,sfms99,binder00&trias00} and vortex
propagation \cite{barahona98}.  In spite various groups have
numerically and theoretically studied the static properties of
Josephson ladders~\cite{barahona98,grimaldi96}, so far no systematic
comparison with experimental data has been carried out.  Grimaldi et
al.\cite{grimaldi96} have performed numerical simulations on these
systems.  Their findings are that the behavior of ladders is quite
different from that of one-dimensional (1D) parallel arrays.  In
contrast to a ladder, a 1D parallel array contains only Josephson
junctions in the direction of the bias current $I_{\rm B}$ but no one
transverse to it.  A 1D parallel array of Josephson junctions placed
in magnetic field shows a pattern of critical current with as many
minima as the number of loops (for relatively low $\beta_{\rm L}$, as
defined below) \cite{miller91,experiments}. In Josephson ladders with
junctions on the horizontal branches not only the number of minima in
the pattern does not correspond to the number of loops even for
extremely low $\beta_{\rm L}$, but also the pattern dependence on the
parameter $\beta_{\rm L}$ is different: the critical current $I_{\rm
  C}$ never reduces to zero for fully frustrated arrays (i.e.\ when
there is half flux quantum in each cell).  Baharona et
al.~\cite{barahona98} have shown that one can analytically estimate
the depinning current of fluxon trapped into the ladder in the limit
of zero inductance. They have also computed the onset of instability
in the case of no fluxons, thus retrieving analytically the numerical
result of Ref.~\onlinecite{grimaldi96} for very low inductance.
Moreover, the authors of Ref.~\onlinecite{barahona98} have estimated
that the critical current of a ladder with a fluxon trapped in
each second cell is higher than the depinning from the empty ground
state for a moderately high magnetic field.

The aim of this paper is to present experiments performed on isotropic
Josephson ladders with various values of the self-inductance parameter
$\beta_{\rm L}$. We call as isotropic a ladder consisting of identical
junctions. We also make an analysis of the model to explain the
observed dependence of the pattern upon $\beta_{\rm L}$.  The work is
organized as follows. In Section II we describe a model for the
Josephson ladders, in Section III we show the experimental findings
and make the comparison with the numerical predictions. Finally,
Section IV contains a discussion of our results and Section V the
conclusion.

\section{The model}

To derive the equations for the ladder, we start from the fluxoid
quantization over a cell:
\begin{equation}
\sum_{cell} \varphi = \frac{2 \pi}{\Phi_0} \left[ \Phi^{\rm ext} +
\Phi^{\rm ind} \right],
\end{equation}
\noindent where $\varphi$ are the phases of all the junctions in the cell,
$\Phi_0$ is the flux quantum, and $\Phi^{\rm ext}$ and $\Phi^{\rm
  ind}$ are the applied and induced flux, respectively.  To evaluate
$\Phi^{\rm ind}$ we retain only self-inductance terms, i.e.\ we assume
that $\Phi^{\rm ind} = L I^s$, $I^s$ being the screening current
circulating in the elementary cell and $L$ is the self inductance of
the cell.  For the junctions we assume the $RCSJ$ model and suppose
that all junctions are identical (isotropic ladder).  With these
ingredients it is possible to derive the following set of equations
for the gauge invariant phase difference across the vertical
($\phi_i$) and horizontal ($\psi_i$) junctions \cite{caputo98}:

\begin{eqnarray}
\label{Genequation}
&\ddot{\phi_i} + \alpha \dot{\phi_i} + \sin{\phi_i}=&\nonumber\\
\lefteqn{=\frac{1}{\beta_{\rm L}} \left[\phi_{i-1} - 2\phi_i + \phi_{i+1}
+2 (\psi_i - \psi_{i-1}) \right] + \gamma, \,\,\,\,\, i=2,\dots,N }
~~~~~~~~~~&&\nonumber\\
&\ddot{\psi_i} + \alpha \dot{\psi_i} + \sin{\psi_i}=&\nonumber\\
&=\displaystyle\frac{1}{\eta\beta_{\rm L}} \left[\phi_{i} - \phi_{i+1}
-2 \psi_i  +2\pi f \right], \,\,\,\,\, i=1,\dots,N.&\nonumber\\
\end{eqnarray}
The boundary conditions are:
\begin{eqnarray}
\label{Genequation2}
&\ddot{\phi_1} + \alpha \dot{\phi_1} + \sin{\phi_1} =
\displaystyle\frac{1}{\beta_{\rm L}} \left[\phi_{2} - \phi_1 +2 \psi_1
 + 2\pi f \right] + \gamma ,&\nonumber \\
&\ddot{\phi}_{N+1} + \alpha \dot{\phi}_{N+1} + \sin{\phi_{N+1}} =&\nonumber\\
&=\displaystyle\frac{1}{\beta_{\rm L}} \left[\phi_{N} - \phi_{N+1} 
+2 \psi_{N} - 2 \pi f \right] + \gamma.&
\end{eqnarray}
Here, $\gamma = I_{\rm B}/i_{\rm c}^{\rm \/ver}$ is the bias current
normalized to the single vertical junction critical current,
$\beta_{\rm L} = 2\pi L i_{\rm c}^{\rm \/ver}/\Phi_0$ is the
self-inductance parameter (the self-inductance $L$ of a square cell
with the side $a$ can be estimated \cite{jaycox81} as $L=1.25 \mu_0
a$, where $\mu_0$ is the magnetic permeability), the ratio
$\eta=i_{\rm c}^{\rm \/hor}/i_{\rm c}^{\rm \/ver}$ between the
horizontal and the vertical junction critical currents is the
anisotropy parameter, $\alpha$ is the normalized dissipation
parameter, $f=\Phi^{\rm ext}/\Phi_0$ is the normalized external flux
often noted as frustration, and $N$ is the number of loops.  For the
static case the parallel arrays considered in
Ref.~\onlinecite{ustinov93,watanabe95,miller91} correspond to the
limit $\eta\rightarrow\infty$. In deriving
Eqs.~(\ref{Genequation})--(\ref{Genequation2}) we take advantage of
the fact that, due to the symmetry of the system, the current flowing
in the top and bottom horizontal junctions of the same cell (see
Fig.~\ref{scheme}(a)) differs only in direction but not in amplitude,
and therefore we can write the equation for just one of them
\cite{filatrella95}.  Finally, we want to stress that in this work we
are interested only in the transition point from the static to the
dynamic solutions (the critical current), therefore the dynamics is
nothing but a computational mean to find the current point at which
the static solution becomes unstable. The value of the dissipation
used in the simulations is fictitious, and it has been chosen equal to
$1$ for computational convenience. The experimentally studied arrays
are actually underdamped and therefore have a much smaller damping
coefficient, $\alpha \simeq 0.005$ -- $0.08$, that can be controlled
by temperature.

\section{Experimental $I_{\rm C}$ vs $\lowercase{f}$ patterns}

We present an experimental study of 10-cell Josephson junction ladders.
Each elementary cell of the ladders contains 4 identical small
Nb/Al-AlO$_x$/Nb Josephson tunnel junctions\cite{Hypres}, which have
an area of $3\times 3 \, \mu \rm m^2$.  To get different values of
$\beta_{\rm L}$, we used samples with different critical current
density $j_{\rm c}$ ($100 \, {\rm A/cm}^2$ or $1000 \, {\rm A/cm}^2$)
and also varied the loop size $a$ ($2.8\,\mu \rm m$ or $9.9\,\mu \rm
m$).  A SEM image of a typical ladder is shown in Fig.~\ref{scheme}(b).

We have measured the ladder critical current, $I_{\rm C}$, versus
frustration $f$ for $4$ isotropic ladders ($\eta=1$) with different
$\beta_{\rm L}$.  The values selected are $\beta_{\rm L} = 3$, $0.88$,
$0.25$ and $0.088$, a range where the peculiar behavior of the ladders
should be clearly visible.  The measurements were performed in a
cryoperm shield.  The magnetic field $H$ applied perpendicular to the
substrate was provided by a coil placed inside the shield. The uniform
bias current $I_{\rm B}$ was injected at every node via on-chip
resistors.  The voltage across the first vertical junction was
measured to define the depinning current of the ladder.  Finally, the
$I_{\rm C}$ vs $f$ dependencies were measured using GoldExi
software\cite{goldexi}.

\begin{figure}
\vspace{5pt}
\centerline{\epsfig{file=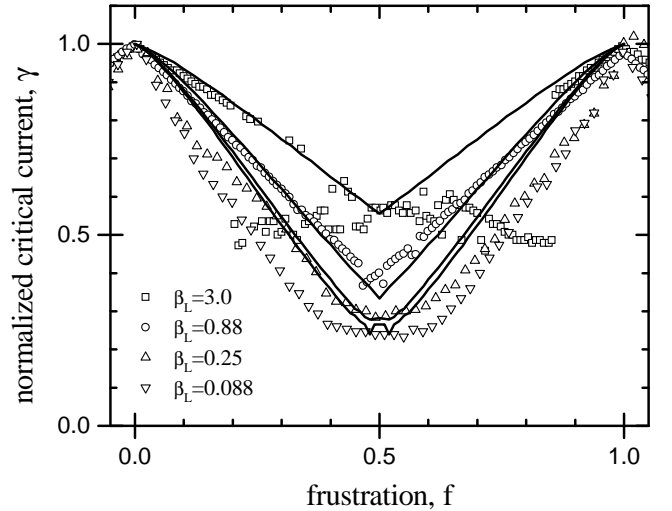,width=3.4in}}
\vspace{5pt}
\caption{Experimental (symbols) and numerical (solid lines) $\gamma$ vs $f$
  patterns of ladders with different cell sizes.  Parameters are: $N =
  11$; $\beta_{\rm L} =3.0$ (squares), $0.88$ (circles), $0.25$ (up
  triangles), $0.088$ (down triangles).}
\label{pattern}
\end{figure}

In Fig.~\ref{pattern} we present two measured features of the ladders.
In contrast to the case of a 1D parallel array, there are no
additional lobes between $f=0$ and $f=1$.  Also, the critical current
$I_{\rm C}$ remains relatively large, despite a $\beta_{\rm L}$ as low
as $0.088$.  Similarly to Ref.~\onlinecite{grimaldi96} we numerically
solved the Eqs.~(\ref{Genequation})--(\ref{Genequation2}) by using the
same parameters as in the experiment (except for $\alpha$, see above).
In Fig.~\ref{pattern}, we compare the numerical simulations (solid
lines) with the experimental data (symbols), which show good
agreement. The $\beta_{\rm L}$ used in the simulations was calculated
from the critical current of a single junction measured in the
experiment.  The calculations show some flattening at $f=0.5$ for low
$\beta_{\rm L}$, which is also present in the experimental data.  We
found in simulations that in this region the ladder gets first filled
with flux and only subsequently undergoes the depinning.  We have
observed this behavior only for low inductance, in good agreement with
the analytical prediction of Ref.~\onlinecite{barahona98} (that
neglects inductance terms).  In experiments with $\beta_{\rm L} = 3$
and $\beta_{\rm L} = 0.88$ we note the simultaneous presence of two
different states at the same frustration value.  We suppose that this
is due to distinctly different initial conditions that can be realized
in the ladder, while sweeping the bias current $I_{\rm B}$ through
zero.  This contradicts the prediction of Ref.~\onlinecite{barahona98}
where it is stated that the depinning of the whole ladder from a state
different from the empty ground state occurs {\em only\/} around
$f=0.5$.  The reason of this disagreement might be the neglect of the
inductance in their calculations.

\begin{figure}
\vspace{5pt}
\centerline{\epsfig{file=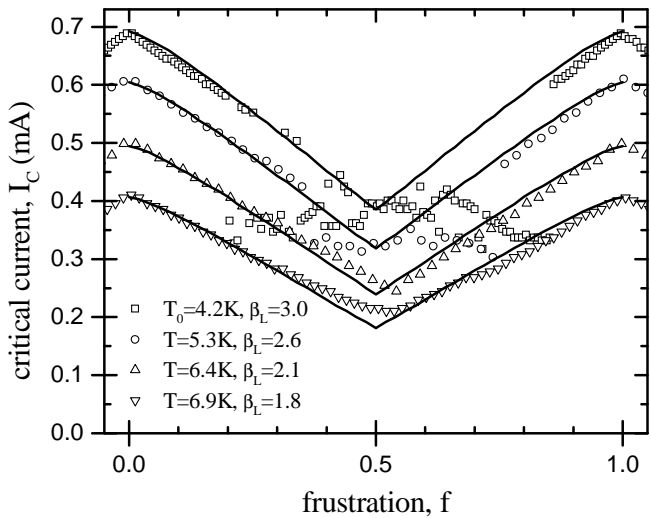,width=3.4in}}
\vspace{5pt}
\caption{Experimental (symbols) and numerical (solid lines) $I_{\rm
    C}$ vs $f$ patterns of one of the ladders measured at different
  temperatures, in order to vary the critical current.  The
  temperature has been derived from the gap voltage.  Parameters are:
  $N = 11$; $\beta_{\rm L} =3.0$ (squares), $2.6$ (circles), $2.1$ (up
  triangles), $1.8$ (down triangles).}
\label{temperature}
\end{figure}

In the ladder with the largest $\beta_{\rm L}$ parameter ($\beta_{\rm
  L} = 3.0$), we have measured the $I_{\rm C}$ vs $f$ dependence also
as a function of the temperature $T$. At higher temperatures the
decreased critical current causes a decrease of $\beta_{\rm L}$.  The
results are shown in Fig.~\ref{temperature}, in physical units to
underline the actual change of the critical current. Also in this case
the agreement between the model and the experiments is rather good.

\section{Discussion}

We have characterized the static properties of Josephson ladders for
different values of the self-inductance parameter $\beta_{\rm L}$. The
experimentally observed dependencies of the critical current on
frustration are in good agreement with the numerical simulations and
show that the behavior of the ladder is clearly different from that of
the 1D parallel array (see Fig.~\ref{pattern}).  As it is well-known,
in 1D parallel arrays the critical (depinning) current is determined
by the parameter $\beta_{\rm L}$, and in the limit of small
$\beta_{\rm L}$ the minimum frustration-dependent critical current is
very small.  Instead, the ladder critical current, even in the case of
small $\beta_{\rm L}$, never goes to zero. As it was already pointed
out in Ref.\onlinecite{grimaldi96}, with respect to 1D parallel
arrays, the presence of the horizontal junctions in the ladder leads
to an "effective" increased $\beta_{\rm L}^{\rm \/eff}$, which for
small discreteness can be by up to two orders of magnitude larger than
the {\em natural\/} $\beta_{\rm L}$ of the system, calculated (similar
to 1D arrays) from the junction critical current and the cell
inductance.

In order to show the particular mapping between Josephson ladders and
1D parallel arrays, we carry out a simple quantitative analysis of the
Eqs.~(\ref{Genequation}). Let us consider the static case, when all
Josephson junction phases are independent of time and satisfy the
system of nonlinear equations:
\begin{equation}
\label{PhaseConn}
\eta(\sin \psi_{i-1}-\sin \psi_{i})~=~\sin \phi_i-\gamma,~~~~~~ i=2, N.
\end{equation}
By making use of the particular assumption that the horizontal phases
$\psi_i$ are small, we can eliminate the phases $\psi_i$ from all
equations and write the system of equations for phases $\phi_i$ in the
form:
\begin{equation}
\label{VertPhase}
\sin \phi_i ~=~ \frac{\eta}{\eta\beta_{\rm L}+2}
\left[\phi_{i-1} - 2\phi_i + \phi_{i+1}
 \right] + \gamma
\end{equation}
This system of equations coincides with the one describing the static
properties of 1D parallel array (with horizontal
junctions replaced by superconducting electrodes). The difference
between a ladder and 1D parallel array is that for the
ladder we have now to use an effective parameter
\begin{equation}
\label{Mapp}
\beta_{\rm L}^{\rm \/eff}=\beta_{\rm L}+2/\eta.
\end{equation}
The deviation of $\beta_{\rm L}^{\rm \/eff}$ from $\beta_{\rm L}$
originates from an additional shielding (and vortex pinning) due to
the presence of horizontal junctions, i.e.\ the horizontal junctions
can accommodate part of the phase change. Thus, we expect that
this deviation disappears in anisotropic ladders when the critical
current of horizontal junctions $i_{\rm c}^{\rm \/hor}$ is much larger
than the critical current of vertical junction $i_{\rm c}^{\rm \/ver}$
($\eta\gg1$).

\begin{figure}
\vspace{5pt}
\centerline{\epsfig{file=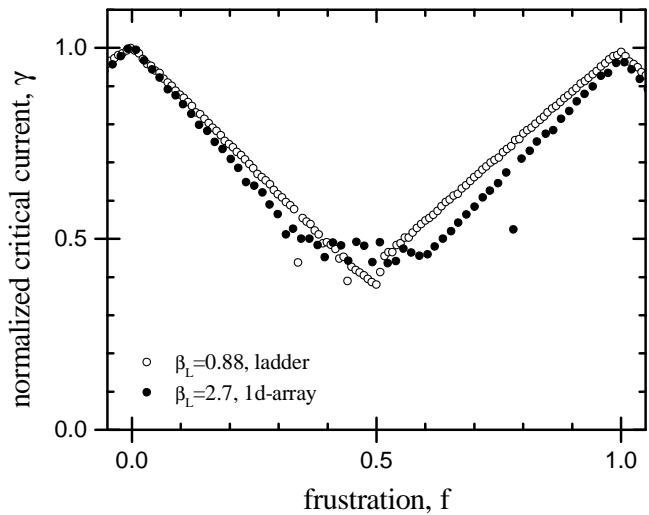,width=3.4in}}
\vspace{5pt}
\caption{$I_{\rm C}$ vs $f$ dependencies of the 1D parallel 
  array with $\beta_{\rm L} \,=\,2.7$ (solid circles) and the ladder
  with $\beta_{\rm L} =0.88$ (open circles).}
\label{ladder-array}
\end{figure}

To verify the mapping given by Eq.(\ref{Mapp}), in
Fig.~\ref{ladder-array} we compare the patterns of a 1D parallel array
with $\beta_{\rm L}^{\rm \/1D} = 2.7$ and an isotropic ladder with
$\beta_{\rm L} = 0.88$.  For the ladder we expect $\beta_{\rm L}^{\rm
  \/eff}\approx\beta_{\rm L}^{\rm \/1D}$.  The agreement is
particularly good at low frustration, but not in the vicinity of
$f\,=\,0.5$, where the most critical assumption of our theory, i.e.\ 
is small values of the horizontal junction phases, breaks down.

We would like to note here that a similar analysis for anisotropic
ladders in the limit of small $\beta_{\rm L}$ has been carried out in
Ref.\onlinecite{Stroud}.  Moreover, for the case of ladder with three
junctions per cell the mapping takes the form: $\beta_{\rm L}^{\rm
  \/eff}=\beta_{\rm L}+1/\eta$. This mapping is in good accord with
previously published data on the $I_{\rm C}(f)$ dependence for ladder
with three junctions per cell \cite{CaputoApplSuper}.  We want to
stress out here that this mapping is supposed to work only for the
static case.  In the dynamic state, when the Josephson vortices
propagate in the ladder, the phases of the horizontal junctions start
to oscillate and Eqs.~(\ref{PhaseConn}) are not valid anymore,
especially in the regime of large vortex velocities. Theoretical and
experimental investigation of vortex propagation in Josephson ladders
will be reported elsewhere \cite{schuster00}.

\section{Conclusion}
We have reported measurements of the critical current in ladders of
Josephson junctions. The $\beta_{\rm L}$ parameter has been varied by
changing both the geometrical inductance and the critical current of
the junctions.  The results are in good agreement with numerical
simulations, and show a behavior clearly distinct from the case of the
1D parallel Josephson junction array without junctions in the
horizontal branches. Using a simple quantitative analysis, we have
shown that the static properties of 1D parallel arrays and ladders can
be mapped by properly scaling the self-inductance parameter
$\beta_{\rm L}$.  This analysis well agrees with experimental data.

\section*{Acknowledgments}

Financial support from the DAAD Vigoni cooperation program is
acknowledged. P.C.\ and M.V.F.\ thank, respectively, the European
Office of Scientific Research (EOARD) and the Alexander von Humboldt
Stiftung for supporting this work.  The samples were fabricated at
Hypres, Elmsford, New York.



\end{document}